\documentclass[12pt]{article}
\usepackage{epsfig,amsfonts,amssymb}
\usepackage{hyperref}
\usepackage{cite}
\input epsf.sty
\usepackage{cite}

\def\ZZZ{{\hbox{ Z\kern-1.6mm Z}}}
\def\RRR{{\hbox{ R\kern-2.4mm R}}}
\def\CCC{{\hbox{ C\kern-2.0mm C}}}
\def\zzz{{\hbox{z\kern-1mm z}}}

\newcommand{\qeq}{{\hbox{=\kern-2.3mm ? \kern.5mm }}}
\renewcommand{\qeq}{=}

\newcommand{\eps}{\epsilon}

\newcommand{\wJ}{\wt J}

\newcommand{\wt}{\widetilde}
\newcommand{\wh}{\widehat}

\newcommand{\be}{\begin{equation}}
\newcommand{\ee}{\end{equation}}
\newcommand{\ben}{\begin{eqnarray}\displaystyle}
\newcommand{\een}{\end{eqnarray}}

\newcommand{\refb}[1]{(\ref{#1})}
\newcommand{\p}{\partial}
\newcommand{\sectiono}[1]{\section{#1}\setcounter{equation}{0}}

\def\one{{\hbox{ 1\kern-.8mm l}}}
\def\zero{{\hbox{ 0\kern-1.5mm 0}}}

\begin{document}

\baselineskip 24pt

\begin{center}
{\Large \bf Black Hole Hair Removal
}

\end{center}

\vskip .6cm
\medskip

\vspace*{4.0ex}

\baselineskip=18pt

\centerline{\large \rm   Nabamita Banerjee, Ipsita Mandal 
and Ashoke Sen }

\vspace*{4.0ex}

\centerline{\large \it Harish-Chandra Research Institute}

\centerline{\large \it  Chhatnag Road, Jhusi,
Allahabad 211019, INDIA}

\vspace*{1.0ex}
\centerline{E-mail:  nabamita at hri.res.in, ipsita at hri.res.in,
sen at hri.res.in, ashokesen1999 at gmail.com}

\vspace*{5.0ex}

\centerline{\bf Abstract} \bigskip

Macroscopic entropy of an extremal black hole is expected to be 
determined completely by its near horizon geometry. Thus two black
holes with identical near horizon geometries should have identical
macroscopic entropy, and the expected equality between
macroscopic and microscopic entropies will then
imply that they have identical degeneracies of microstates.
An apparent counterexample is provided by the 4D-5D lift relating
BMPV
black hole to a four dimensional black hole. 
The two black holes have identical
near horizon geometries but different microscopic spectrum. We
suggest that this discrepancy can be accounted for by black hole hair,
--  degrees of freedom living outside the horizon and contributing to the
degeneracies. We identify these degrees of freedom for both
the four and the five dimensional black holes and show that after their
contributions are removed from the microscopic degeneracies of the
respective systems, the result for the four and five dimensional black holes
match exactly.  

\vfill \eject

\baselineskip=18pt

\tableofcontents

\sectiono{Introducion and summary}

Since the Bekenstein-Hawking entropy of a black hole is proportional
to the area of the event horizon of the black hole\cite{haw,beck,haw2}
one expects that 
the horizon of the black hole contains the key to understanding the
black hole microstates. Wald's modification of the Bekenstein-Hawking
formula in higher derivative theories of 
gravity\cite{9307038, 9312023, 9403028, 9502009} 
deviates from the area
law, but nevertheless expresses the black hole entropy in terms of the
horizon data. The situation becomes even better in the extremal limit
where an infinite throat separates the horizon from the rest of the black
hole space-time and the near horizon configuration can be regarded as
a fully consistent solution to the field 
equations\cite{9504147,9508072,9602111,9602136, 9812082}.  
The classical
Wald entropy can be related to the value of the classical
Lagrangian density evaluated in this near horizon 
geometry\cite{0506177}.
This leads one to expect that we should be able to define a 
macroscopic quantity, computed from
quantum string theory in the near horizon geometry, that captures
complete information about the microscopic degeneracies of the
corresponding black hole.
Quantum entropy function is such a proposal  relating the microscopic
degeneracies of extremal black holes to an appropriate partition function
of quantum gravity in the near horizon geometry of the black 
hole\cite{0809.3304,0810.3472} 
(see also \cite{0805.0095,0806.0053}). 

Irrespective of any specific proposal,
if the postulate that 
the microscopic degeneracy of an
extremal black hole can be related to some computation in the near
horizon geometry is correct, then this leads to an immediate consequence:
two black holes with identical near horizon geometries will have
identical degeneracies of microstates. There are some trivial 
counterexamples with straightforward resolutions. For example the near
horizon geometry of an extremal black hole in flat space-time
is independent of
the asymptotic values of the moduli fields due to the 
attractor 
mechanism\cite{9508072,9602111,9602136,0506177,0507096,
0606244},
but the microscopic degeneracy of states, carrying the same quantum
numbers as the black hole, jumps across the
walls of marginal stability as we vary the 
asymptotic moduli\cite{0605210,0609109,
0702141,0702150,0706.2363}. The
resolution of this puzzle is provided by the fact that for a given set
of charges there are typically many classical solutions. One of these is
a single centered black hole solution but the others contain multiple
centers\cite{0005049, 0010222, 0101135, 0206072,
0304094, 0702146,0705.3874,0706.2363}. 
As we cross a wall of marginal stability some of these
multi-centered solutions cease to exist and hence cause 
a jump in the
total entropy. This precisely accounts for the jump in the
total degeneracy across the walls of marginal stability, thereby
showing that the degeneracy of states associated with
a single centered 
black hole remains unchanged as we cross a wall of marginal stability.
This suggests a natural modification of the original proposal: {\it
string
theory in the
near horizon geometry captures information about the
microscopic degeneracy of the
single centered black holes only.} This is clearly natural from a
physical perspective: the near horizon geometry of a given black
hole should encode information only about the particular solution 
which produces the particular near horizon geometry. Multi-centered
black holes have multiple horizons with multiple near horizon geometry,
and hence the contribution to their degeneracies should involve
studying string theory in
the near horizon geometry of each of these black 
holes.\footnote{The near horizon $AdS_2$ geometry of a black hole
can fragment into multiple throats carrying different 
charges\cite{9202037,9812073,0504221}. 
However for such solutions the charges carried by the
fragments are mutually local, \i.e.\ have 
$(\vec q_i\cdot \vec p_j-\vec q_j\cdot \vec p_i)=0$ 
where $(\vec q_i, \vec p_i)$ denote the electric and magnetic charge
vectors of the $i$th throat.
Since such configurations do not
contribute to the entropy\cite{0206072,0702146},
the conclusion that
the near horizon geometry of a black hole captures the degeneracies
of single centered black holes remains unchanged.}

In order to make this modified proposal concrete we 
must independently
define microscopic degeneracy of a single centered black hole. 
Typically
microscopic computation involves studying 
degeneracies of
various brane configurations and cannot distinguish whether a given
state would correspond to a single centered or a multi-centered 
configuration in the limit when the state becomes a black hole.
However in asymptotically  flat four dimensional space-time
there is  a simple algorithm for calculating the spectrum of
single centered black holes in the microscopic theory; we simply need
to set the asymptotic values of the moduli to be equal to their attractor
values.\footnote{This is sufficient but not necessary; all we need is
that the asymptotic values of the moduli should be chosen such that we
can continuously deform them to the attractor values without 
crossing any
wall of marginal stability.}
In that case all multi-centered black hole solutions disappear
and the microstate counting  
only picks up the contribution from the single centered black holes.

In this paper we focus on a different counterexample that cannot be
resolved by invoking the existence of multi-centered black holes.
This involves the BMPV black 
hole\cite{9602065},
whose microscopic description involves a D1-D5 
system of type IIB string
theory on $K3\times S^1$, carrying momentum along $S^1$ 
and equal angular momentum in two planes transverse to the D5-brane. 
The macroscopic description
of this is a five dimensional rotating black hole. By placing this black
hole at the center of a Taub-NUT space we get a four dimensional
black hole\cite{0209114}. Since near the 
origin the Taub-NUT space appears as
flat space, the near horizon geometries of the four and five dimensional
black holes are exactly identical\cite{0503217,0505094}. 
However the 
microscopic description
of the four dimensional black hole involves D1-D5-brane moving in the
background of a Kaluza-Klein monopole and the degeneracies of 
this system are different from those of
just the D1-D5 system\cite{0605210}. 
This would seem to
contradict the claim that the microscopic degeneracies of
single centered black holes are completely encoded in their near horizon
geometries.

We suggest the following resolution of this puzzle. Common sense 
tells us that the near horizon
geometry should capture the degeneracies associated with the
dynamics of the horizon. If the black hole has no hair, that is
no degree of freedom living outside the horizon that could
contribute to the degeneracy, then the near horizon geometry would
capture the complete information about the microscopic degeneracy
of the black hole. However
if the black hole solution contains degrees of freedom living outside the
horizon then the full degeneracy of the black hole has to be computed
by combining the contribution from the horizon with
the contribution from the degrees of freedom living outside the horizon,
and the combined contribution will then have to be compared with the
microscopic degeneracies. 
Thus two black holes having identical near
horizon geometry can have different microscopic degeneracies if they
have different sets of degrees of freedom living outside the horizon.
We expect that at least
for extremal black holes the separation 
between the contribution from
the black hole hair and the contribution from the horizon degrees
of freedom can be done rigorously since the horizon is separated
from the asymptotic space-time by an infinite throat. Thus two such
extremal black holes with identical near horizon geometry will have
identical degeneracies of microstates {\it after we remove the
contribution from the degrees of freedom living outside the
horizon.}\footnote{This is similar in spirit to the phenomenon that
for a stack of $N$ D3-branes, string theory living in the bulk of the
near horizon $AdS_5\times S^5$ geometry does not capture the
$U(1)$ center of mass degrees of freedom of the 
D3-branes\cite{9711200}.}
 
In the rest of the paper we shall identify the degrees of freedom living
outside the horizon for both the BMPV black hole and the
four dimensional extremal black hole obtained by placing the
BMPV black hole in a Taub-NUT geometry, 
and then show that their microscopic degeneracies agree
after we remove the contribution 
due to the hair. The organisation of the sections will be as follows.
In \S\ref{s2} we identify the hair degrees of freedom of the five
dimensional BMPV black hole, and remove their contribution
from the partition function to determine the partition function 
associated with the horizon degrees of freedom. The result is
given in \refb{ese2}. In \S\ref{s3} we repeat the same analysis
for the four dimensional black hole obtained by placing the 
BMPV black
hole at the center of Taub-NUT space. The result, given in
\refb{eadd2}, is found to agree with \refb{ese2}. It of course
remains a challenge to reproduce these microscopic results from
a macroscopic calculation, {\it e.g.} of the quantum entropy
function.
In the two appendices we describe explicit construction of the
bosonic modes associated with the hair degrees of freedom.

Before concluding this section we would like to add a word of caution.
While we have identified appropriate hair degrees of freedom for the
five and four dimensional black holes
after whose removal the result for the partition function of the two
black holes agree, we have not proved that these are the 
only hair degrees
of freedom. If there are additional hair degrees of freedom which differ
for these two black holes then it could spoil the agreement. On the
other hand if there are additional hair degrees of freedom which are
common to both black holes then the agreement between the partition
functions of the two black holes after hair removal will continue
to hold.

\sectiono{Analysis of the BMPV black hole entropy} \label{s2}

We begin with the analysis
of microscopic degeneracy of the five
dimensional quarter BPS black hole in type IIB string theory on
$K3$. The microscopic description
involves $Q_5$  number of D5-branes
wrapped on $K3\times S^1$ and $Q_1$  
number of D1-branes wrapped
on $S^1$ carrying $-n$ units of momentum along $S^1$ 
(with $n> 0$) and
 $J$  units of angular momentum. 
For
simplicity we 
shall take $Q_5=1$ without any loss of generality since the result depends
on $Q_1$ and $Q_5$ only through the combination 
$Q_5(Q_1-Q_5)$. 
Our convention for angular momentum and supersymmetry
generators will be as follows. 
We denote
the $SO(4)$ rotation group of the five dimensional space-time
by $SU(2)_L\times SU(2)_R$ and identify 
the angular momentum $J$ with
twice the diagonal generator of $SU(2)_L$. We also denote by
$h$ the eigenvalue of the diagonal generator of $SU(2)_R$.
Since supersymmetry transformation parameters of type IIB on
K3 are chiral spinors in six dimensions, when we regard them as
representations of the $SO(1,1)\times 
SU(2)_L\times SU(2)_R$ subgroup of the
Lorentz group, with $SO(1,1)$ acting on the common direction of
the D1-brane and the D5-brane, the $SO(1,1)$ quantum numbers
will be correlated with the $SU(2)_L\times SU(2)_R$ 
quantum numbers. 
We shall now argue that in order that the configuration described
above describes a quarter BPS state, we must 
choose the convention that the left-chiral spinors  of
$SO(1,1)$ carry $(J=0, 2h=\pm 1)$ and the right-chiral spinors
of $SO(1,1)$ carry $(J=\pm 1, h=0)$. 
The argument goes as follows. First of all note that since the
D1-D5-brane system carries negative momentum along $S^1$,
it must be allowed to carry left-moving excitations without
violating supersymmetry. Thus the left-chiral excitations must
be neutral under the unbroken supersymmetries of the
system. This in turn implies that these supersymmetry
transformation parameters must be left-chiral spinors of
$SO(1,1)$, -- since left-chiral supersymmetry transformation
parameters act on the right-chiral
modes and vice versa. 
We shall now argue that the unbroken 
supersymmetry transformation
parameters must also carry $J=0$, -- this would force us to
choose the convention described above.
In order that the system can carry
macroscopic $J$ charge, a large number of internal modes must
carry non-vanishing $J$ charge. Now most of the bosonic
degrees of
freedom come from the motion of the D1-brane inside the D5-brane,
\i.e.\ along the $K3$ direction. 
This leads to four bosons for each D1-brane describing its position
along $K3$.
These modes
are clearly
neutral under the $SO(4)$ rotation along the space transverse
to the D1-D5-brane  system, and hence
do not carry any $J$ charge. On the other
hand for every D1-brane we also have eight fermionic modes,
-- four carrying $(J\ne 0, h=0)$ and four carrying 
$(J=0, h\ne 0)$.\footnote{These have opposite relation between
the $SO(1,1)$ and $SU(2)_L\times SU(2)_R$ quantum numbers,
but we shall not need to use this information here.}
The requirement of unbroken supersymmetry
freezes  the modes on which supersymmetry acts, \i.e.\ those which
form partners of the bosons. Now since we want to excite the
modes carrying $J$ charge, we must freeze the ones with $J=0$.
Thus the latter must be acted upon by supersymmetry and paired
with the bosons. Since  the bosons carry $J=0$, the supersymmetry
transformation parameter must also carry $J=0$. This establishes the
desired result.

We denote by
$d_{5D}(n,Q_1,J)$ the helicity trace 
$-Tr\left((-1)^{2h+J}\, (2h)^2\right)/2!$ of
five dimensional black hole carrying 
quantum numbers $(n,Q_1,J)$, and
define
\be \label{ep1}
Z_{5D}(\rho,\sigma,v) = \sum_{n,Q_1,J} \, d_{5D}(n,Q_1, J)\, 
\exp\left[2\pi i \{(Q_1-1) \,  \sigma + (n-1) \,  \rho + J v\}\right]\, .
\ee  
The $-1$ in $(Q_1-1)$ reflects the
fact that  a D5-brane wrapped on $K3$ carries $-1$ units of
D1-brane charge. On the other hand the $-1$ in $(n-1)$ has been 
introduced due to the fact that this charge measured at $\infty$
differs from that measured on the 
horizon\cite{0705.1847,0807.0237,0807.2246}
-- a Chern-Simons coupling
in the action produces $-1$ unit of this charge from the region between
the horizon and infinity. Thus if $-n$ is the total momentum along $S^1$
carried by the black hole, the charge measured at the horizon will be
$-(n-1)$.
Explicit computation shows that 
$Z_{5D}$ defined in \refb{ep1} has
the form
\ben\label{ep2}
Z_{5D}(\rho,\sigma,v) &=&
e^{-2\pi i  
\rho-2\pi i \sigma
 }\prod_{k,l,j\in \zzz 
\atop k\ge 1, l\ge 0}
\left( 1 - e^{2\pi i ( \sigma k   +  \rho l + v j)}
\right)^{-c(4lk - j^2)} \nonumber \\
&& \times \left\{\prod_{l\ge 1} (1-e^{2\pi i(l\rho+v)})^{-2} \, 
(1-e^{2\pi i(l\rho-v)})^{-2} \, (1-e^{2\pi il\rho})^4 \right\}\,
(-1)\, (e^{\pi i v}-
e^{-\pi i v})^2 \nonumber \\
&& + \, e^{-2\pi i  
\rho-2\pi i \sigma
 } \, \prod_{k, j\in \zzz 
\atop k\ge 1}
\left( 1 - e^{2\pi i ( \sigma k   + v j)}
\right)^{-c(  - j^2)} \,  (e^{\pi i v}-
e^{-\pi i v})^2\, , \een
where the coefficients $c(n)$ are defined via the equation
\be \label{ep3}
8\, \left[ {\vartheta_2(\tau,z)^2
\over \vartheta_2(\tau,0)^2} +
{\vartheta_3(\tau,z)^2\over \vartheta_3(\tau,0)^2}
+ {\vartheta_4(\tau,z)^2\over \vartheta_4(\tau,0)^2}\right]
= \sum_{j,n\in\zzz} c(4n - j^2) \, e^{2\pi i n\tau + 2\pi i j z}\, .
\ee
Eq.\refb{ep2} requires some explanation. The first line of
\refb{ep2} denotes the contribution from the relative motion of the
D1-D5 system and was computed in \cite{9608096}. 
The asymptotic expansion of the degeneracies of this system has been
studied recently in \cite{0807.0237,0807.1314}.
The second line
represents contribution from the `center of mass modes' 
of the D1-D5
system.  This contribution can be calculated as follows.
Since the D1-D5 system breaks the translation symmetries along the
four directions transverse to the brane, the (1+1) dimensional
world-volume theory
of this system,  spanned by the time coordinate and 
the coordinate along $S^1$, will contain
four goldstone bosons 
associated with the four broken translation generators. Furthermore
since the ground state of the D1-D5 system also breaks eight out
of the sixteen supersymmetries of type IIB string theory on K3, 
we shall
have eight goldstino fermions carrying the same quantum numbers
as the broken supersymmetry transformation parameters. 
This leads to four left-moving and four
right-moving fermions living on the D1-D5-brane 
world-volume.  In our
convention the left-moving fermions carry
$(J=0,2h=\pm 1)$ and the right-moving fermions carry 
$(J=\pm 1, 2h=0)$. 
We need to count excitations of this system
preserving four  supersymmetries,
parametrized by left-chiral spinors
on the D1-D5-brane world-volume. 
Since these transformations act on the right-moving
fermions and bosons, 
the BPS condition will freeze all the right-moving 
excitations except the zero modes.
Since the right-moving fermions carry $J=\pm 1$, $h=0$,
quantization of a pair of right chiral zero modes would produce
a pair of states with $J=\pm{1\over 2}$, $h=0$.
Thus the net contribution of four right chiral
zero modes
to the trace, containing a factor of 
$(-1)^J e^{2\pi i v J}=e^{2\pi i J (v+{1\over 2})}$,
is a factor of $(e^{\pi i (v+{1\over 2})}+
e^{-\pi i (v+{1\over 2})})^2=-(e^{\pi i v}-
e^{-\pi i v})^2$. This accounts for the last two factors in 
the second line of \refb{ep2}.
The BPS condition does not restrict the
left-moving degrees of freedom and
the terms in the curly bracket in
the second line of \refb{ep2} represent contribution from these
left-moving excitations. In particular 
the zero modes of the 
left-moving fermions, carrying helicities $\pm 1/2$, soak up the
factors of $-(2\, h)^2/2!$ 
in the helicity trace so that if we leave aside these
zero modes, contribution to the helicity trace from 
the rest of the modes involve computing the Witten index
$Tr(-1)^F$. Since the left-moving fermions
have $J=0$, their oscillators
lead to the last
term in the product inside the curly bracket. 
On the other hand  the left-moving bosons, transforming
under $(2,2)$ representation of $SU(2)_L\times SU(2)_R$, carry
$\pm1$ units of $J$ quantum numbers and lead to the  first two
terms inside the curly bracket. Finally the term in the last line of
\refb{ep2} removes the contribution of the $n=0$ 
term\footnote{Throughout this 
paper we shall denote the additive term proportional to
$e^{-2\pi i\rho}$ as the $n=0$ term.} from eq.\refb{ep1},
\i.e.\ it subtracts the
term whose $\rho$ dependence is of the form 
$e^{-2\pi i\rho}$.
The rationale for subtracting this term is that for 
$n=0$ the D1-D5 system
includes contribution from 
half-BPS states. Thus it is more natural to consider the partition
function of pure quarter BPS states by subtracting the contribution
due to the $n=0$ term.

Now we need to analyze the contribution to the partition function
from the degrees of freedom of the black hole living outside the
horizon and remove this contribution from \refb{ep2} to
determine the expected microscopic degeneracies associated with the
horizon. 
We begin by writing down the action and the
black hole solution.
The relevant part of the
action containing the string metric $G_{\mu\nu}$, dilaton
$\Phi$ and the Ramond-Ramond 3-form field strength 
$F^{(3)}=dC^{(2)}$ takes the form
\be \label{err1}
{1 \over (2\pi)^7}\,
\int d^{10} x \sqrt{-\det G}\, 
\left[
                e^{-2 \Phi}\left(R + 4\, G^{\mu\nu}\, \p_\mu\Phi
                \p_\nu\Phi\right)
                 - {1 \over12} F^{(3)}_{MNP}F^{(3)MNP}  
                 \right]\, ,
\ee
in $\alpha'=1$ unit.
For simplicity we shall set the asymptotic values of the moduli to their
attractor values for the specific black hole solution
we analyze, so that all the moduli fields including the
dilaton are constants. The generalization 
to more general asymptotic
values is straightforward. In this case 
the rotating black hole solution describing 
$Q_5$ D5-branes along 
$K3\times S^1$, $Q_1$ D1-branes along $S^1$,
$-n$ units of momentum along $S^1$ and angular momentum
$J$, takes the form\footnote{Conventionally 
the BMPV black hole as well
as the BMPV black hole at the center of Taub-NUT space is
expressed as a solution in five dimensional supergravity 
theory\cite{9602065,0209114,0503217}.
Here we express them as solutions in a ten dimensional theory
so that we can study the excitations which propagate along the internal
directions.}
\ben \label{ep4}
dS^2 &=& \left( 1 + { r_0   \over  r}\right)^{-1} 
 \left[ - dt^2 +(dx^5)^2 + {r_0   \over r } (dt +
dx^5)^2 + {\wJ \over 4r} \, (dt+ dx^5) \,
(dx^4 + \cos\theta \, d\phi)  \right]
\nonumber \\
&& +  \wh g_{mn}(\vec u)\,
 du^m du^n 
 + \left( 1 + { r_0   \over  r}\right)\, ds_{flat}^2 \, , \nonumber \\
ds_{flat}^2 &=&    r\, 
(dx^4 +\cos\theta d\phi)^2 +  {1\over r}  \,
(dr^2 + r^2 d\theta^2 + r^2\, \sin^2\theta \, d\phi^2)\, , \nonumber \\
  (\theta,\phi, x^4)&\equiv& (2\pi-\theta, \phi+\pi, x^4+\pi)
\equiv (\theta, \phi+2\pi,x^4+2\pi) \equiv (\theta,\phi,x^4+4\pi)\, ,
\nonumber \\
e^\Phi&=&\lambda \, , \nonumber \\
F^{(3)}&\equiv& 
{1\over 6} \, F^{(3)}_{MNP} dx^M \wedge dx^N\wedge
dx^P \nonumber \\
&=& {r_0\over \lambda}\, 
\left(\epsilon_3 + *_6 \epsilon_3
+{1\over r_0} \left( 1 + { r_0   \over  r}\right)^{-1}
(dx^5+dt)\wedge d \zeta \right)\, , \nonumber \\
\eps_3 &\equiv & \sin\theta\, dx^4 \wedge d\theta\wedge d\phi\, ,
\een
where
$x^5$ is the coordinate of the circle
$S^1$ with period $2\,\pi\, R_5$,
$u^m$ for
$m = 6,...,9$ are the coordinates of $K3$,
$\wh g_{mn}$
is the metric on $K3$,  ${\left( 2\pi \right)^4 V}$ 
is the volume of $K3$
measured in this metric,
$\lambda$ is the asymptotic value of the string coupling,
$*_6$ denotes Hodge
dual in the six dimensions spanned by $t$, $x^5$, $x^4$, 
$r$, $\theta$
and $\phi$ with the convention $\epsilon^{t54r\theta\phi}=1$, and
\begin{equation} \label{ep5}
r_0  = {\lambda (Q_1-Q_5)  
 \over 4V } = {\lambda Q_5  \over 4
}  = { \lambda^2 |n |  \over 4R_5^2 V }\, ,
\end{equation}
\be \label{edefwj}
\wJ = {J\, \lambda^2\over 2\, R_5\, V}\, ,
\ee
\be \label{edefzeta}
\zeta = -{\wJ \over 8r} \, (dx^4 +\cos\theta d\phi)\, .
\ee
Eq.\refb{ep5} determines the asymptotic moduli $V$ and 
$\lambda/R_5^2$
in terms of the charges. This corresponds to 
setting the asymptotic moduli
to their attractor values. 
$ds_{flat}^2$ 
describes flat euclidean space in the Gibbons-Hawking coordinates.
Higher derivative corrections to the entropy of this black hole have
been discussed extensively in 
\cite{0505188,0703087,0703099,0705.1847,0710.3886,
0807.0237,0807.2246,0809.4954}.

Now the black hole solution breaks four
translation symmetries and twelve of the sixteen space-time
supersymmetries, and hence we expect to have four bosonic
zero modes and twelve fermionic zero modes living on the
black hole, forming part of the black hole 
hair.\footnote{Given that black hole solution outside the
horizon changes under these translations and supersymmetry
transformations, it is clear that these modes are
non-vanishing outside
the horizon. What is not apparent at this stage is whether they
have support entirely outside the horizon. For now we shall 
proceed by assuming
that this is the case, but will study this issue in detail 
in appendix \ref{sa}.}
Typically the 
quantization of the bosonic zero modes do not
give rise to additional degeneracies but produces new charge
sectors instead, -- this was illustrated in \cite{0705.1433}
in the context of four dimensional black holes. 
However the quantization of the fermion zero
modes does affect the partition function.
The $(J,h)$ 
quantum numbers of the fermion zero modes can be read out
by comparison with the microscopic description. 
Since the four unbroken supersymmetries are labelled by left-chiral
spinors on the D1-D5 world-volume, eight of the broken supersymmetries
are right-chiral and four of the broken supersymmetries are
left-chiral. This leads to eight right-chiral and four left-chiral
zero modes.
The left-chiral zero modes carrying $(J=0, h=\pm{1\over 2})$ 
soak up the factors of $-(2\, h)^2/2!$ in the helicity
trace, so that for the rest of the degrees of freedom we only
need to calculate the Witten index $Tr(-1)^{2h+J}$.
On the other hand 
the right-chiral zero modes carry 
$(J=\pm 1, h=0)$ and their contribution to the partition function
is given by 
\be \label{ezem}
(e^{\pi i v}-e^{-\pi i v})^4\, .
\ee

This however is not the end of the story.
Given a zero mode we can explore whether it is possible to lift it
to a full fledged field in $(1+1)$ dimensions spanned by the 
coordinates $(t,x^5)$. If we can lift them to such fields then the
oscillation modes of these fields would produce additional
contribution to the partition function of the black hole hair. 
To
this end we note that if the  black hole solution had been Lorentz
invariant in the $(x^5,t)$ plane, then any broken symmetry would
automatically lead to a massless goldstone or goldstino field on the
black hole world volume instead of just the zero modes. 
In particular the bosonic zero
modes would lift to scalar fields, left-chiral fermion zero modes would
lift to left-moving fermion fields and right-chiral fermion zero modes
would lift to right-moving fermion fields.
However the black hole solution \refb{ep4} does not have
$(1+1)$ dimensional Lorentz invariance, and hence {\it a priori}
we cannot use results in 1+1 dimensonal quantum field theory to
conclude that associated with a broken symmetry we shall have a
massless field living on the world-volume of the 
black hole. Nevertheless we shall now argue that 
the left-moving modes are not affected by the breaking of Lorentz
invariance and continue to exist. Our argument will be somewhat
heuristic, but we compensate for it by giving a detailed construction of
these modes in appendix \ref{sa}. First we
note that the source of Lorentz non-invariance in \refb{ep4} are
the
$(dt+dx^5)^2$ term and the $(dt+ dx^5) \,
(dx^4 + \cos\theta \, d\phi)$ terms
in the metric. 
This structure of the metric shows that only the
$g_{++}$ and $g_{+i}$ 
components of the metric violate the Lorentz invariance.
Since these lead to $g^{--}$ and $g^{-i}$ components of the 
metric but no $g^{++}$ or $g^{+i}$ components, we
see that the Lorentz violating terms in the equation of motion of
various modes around the solution must involve $\p_-$ derivatives
or $_{-\cdots}$ components of fields. In particular the
left-moving fields $\varphi$
for which $\p_- \varphi=0$ do not couple to the $g^{--}$ or
$g^{-i}$ components of the metric
and should continue to describe solutions to linearized equations
of motion around the black hole background.
Thus we can conclude that the world-volume
of the black hole
will have four left-moving bosonic fields carrying
$(J=\pm 1, 2h=\pm 1)$ and 
four left-moving fermion fields carrying $(J=0, 2h=\pm 1)$.
Their contribution to the
partition function is given by
\be \label{einter}
\prod_{l\ge 1} (1-e^{2\pi i(l\rho+v)})^{-2} \, 
(1-e^{2\pi i(l\rho-v)})^{-2} \, (1-e^{2\pi il\rho})^4\, .
\ee
Multiplying this by the contribution \refb{ezem}
from the zero modes we get the
total
contribution to the partition function from the degrees of freedom
living outside the horizon
\be \label{ese1}
Z^{hair}_{5D}(\rho,\sigma,v)
=  (e^{\pi i v}-e^{-\pi i v})^4\,
\prod_{l\ge 1} (1-e^{2\pi i(l\rho+v)})^{-2} \, 
(1-e^{2\pi i(l\rho-v)})^{-2} \, (1-e^{2\pi il\rho})^4 \, .
\ee

Let $Z_{5D}^{hor}(\rho,\sigma,v)$ denote the partition
function
associated with the horizon degrees of freedom of the five dimensional
black hole. Naively we have the relation $Z_{5D}=Z_{5D}^{hor}
\times Z_{5D}^{hair}$. However we shall now argue that there
is an extra additive contribution to $Z_{5D}$, and the correct relation
is
\be \label{ecor1}
Z_{5D} = Z_{5D}^{hor}
\times Z_{5D}^{hair} + Z_{5D}^{extra}\, .
\ee
The extra contribution $Z_{5D}^{extra}$ comes from
starting with a configuration where the black hole does not carry any
momentum along $S^1$, and then exciting its hair degrees of freedom
carrying momentum. As can be seen from \refb{ep5},
the initial configuration is singular in the supergravity
approximation. Thus it describes
a `small black hole' in five dimensions,\footnote{Here, as well as in
\S\ref{s3}, we shall denote by `small black hole' any object which
is singular in the supergravity limit, 
carrying $Q_1$, $Q_5$ and $J$ quantum numbers
but no momentum along $S^1$. Thus it includes  small black
ring configurations as well\cite{0506215,0611166}.}
and hence
its hair degrees of freedom are different from the ones we analyzed
earlier. In particular since the D1-D5 system without momentum
breaks only four left-chiral and four right chiral supersymmetries, 
we have only four right chiral zero modes instead of 8, and hence
a factor of $-(e^{\pi i v}-e^{-\pi i v})^2$ will be missing from the
hair degrees of freedom. Furthermore since the D1-D5-brane
world-volume theory now has full (1+1) dimensional Lorentz
invariance, the right-chiral modes are now
lifted to full right-moving fields, However the requirement of unbroken
supersymmetry still
freezes the right-moving excitations to their ground
state. 
Thus the net contribution from the hair is given by
\be \label{ecor2}
Z_{small}^{hair}=-(e^{\pi i v}-e^{-\pi i v})^2\,
\prod_{l\ge 1} (1-e^{2\pi i(l\rho+v)})^{-2} \, 
(1-e^{2\pi i(l\rho-v)})^{-2} \, (1-e^{2\pi il\rho})^4 \, .
\ee
Let us denote by $Z_{small}^{hor}$ the contribution from the
horizon degrees of freedom of the small black hole. 
Then $Z_{5D}^{extra}$ will be obtained by taking the product
$Z_{small}^{hor}\times Z_{small}^{hair}$ 
and subtracting the $n=0$
contribution. 
On the other hand $Z_{small}^{hor}$ may be determined by
identifying the $n=0$ contribution in 
$Z_{small}^{hor}\times Z_{small}^{hair}$ 
with the partition function of the 
D1-D5 system with no momentum along $S^1$. The latter is
simply the negative of the last term in \refb{ep2}:
\be \label{ecor3}
-e^{-2\pi i  
\rho-2\pi i \sigma
 } \, \prod_{k, j\in \zzz 
\atop k\ge 1}
\left( 1 - e^{2\pi i ( \sigma k   + v j)}
\right)^{-c(  - j^2)} \,  (e^{\pi i v}-
e^{-\pi i v})^2\, .
\ee
Dividing \refb{ecor3} by the $\rho$ independent term in the
series expansion of
\refb{ecor2} gives
\be \label{ecor4}
Z_{small}^{hor}(\rho,\sigma,v) =
e^{-2\pi i  
\rho-2\pi i \sigma
 }\, \prod_{k, j\in \zzz 
\atop k\ge 1}
\left( 1 - e^{2\pi i ( \sigma k   + v j)}
\right)^{-c(  - j^2)} \, .
\ee
$Z_{5D}^{extra}$ is now obtained by multiplying \refb{ecor4}
by \refb{ecor2} and then subtracting the $n=0$ term, \i.e.\
the term proportional
to $e^{-2\pi i\rho}$ in the
series expansion:
\ben \label{ecor5}
Z_{5D}^{extra}(\rho,\sigma,v) &=& - e^{-2\pi i  
\rho-2\pi i \sigma}\,  (e^{\pi i v}-e^{-\pi i v})^2\, 
\prod_{k, j\in \zzz 
\atop k\ge 1}
\left( 1 - e^{2\pi i ( \sigma k   + v j)}
\right)^{-c(  - j^2)} \nonumber \\ &&
\times \prod_{l\ge 1} (1-e^{2\pi i(l\rho+v)})^{-2} \, 
(1-e^{2\pi i(l\rho-v)})^{-2} \, (1-e^{2\pi il\rho})^4 \nonumber \\
&& + e^{-2\pi i  
\rho-2\pi i \sigma
 } \,  (e^{\pi i v}-
e^{-\pi i v})^2 \, \prod_{k, j\in \zzz 
\atop k\ge 1}
\left( 1 - e^{2\pi i ( \sigma k   + v j)}
\right)^{-c(  - j^2)} \, .
\een
Using \refb{ep2}, \refb{ese1}, \refb{ecor1} and \refb{ecor5} 
we now get
\ben \label{ese2}
Z_{5D}^{hor}(\rho,\sigma,v) &= & (Z_{5D}-Z_{5D}^{extra})
/Z_{5D}^{hair} \nonumber \\
&=& -e^{-2\pi i  
\rho-2\pi i \sigma
 }\, (e^{\pi i v}-e^{-\pi i v})^{-2}\,
 \prod_{k,l,j\in \zzz 
\atop k\ge 1, l\ge 0}
\left( 1 - e^{2\pi i ( \sigma k   +  \rho l + v j)}
\right)^{-c(4lk - j^2)}\nonumber \\
&& +  e^{-2\pi i  
\rho-2\pi i \sigma
 }\, (e^{\pi i v}-e^{-\pi i v})^{-2} \, \prod_{k, j\in \zzz 
\atop k\ge 1}
\left( 1 - e^{2\pi i ( \sigma k   + v j)}
\right)^{-c(  - j^2)}  \, .
\een
The presence of the $(e^{\pi i v}-
e^{-\pi i v})^{-2}$
factor may lead one to believe that $Z_{5D}^{hor}$ has a double
pole at $v=0$ and hence the index extracted from this partition
function will suffer from the contour prescription ambiguities
discussed in \cite{0702141,0702150,0706.2363}. However using
the relation $\sum_j\, c(4n-j^2)=24\, \delta_{n,0}$ and the
$v\to-v$ symmetry one can show that the sum of the two terms
in \refb{ese2} has no singularity at $v=0$.
Thus \refb{ese2} leads to an
unambiguous result for the index of quarter BPS states associated with
the horizon degrees of freedom.
We also note that since the factor of $-(2\, h)^2/2!$ 
in the helicity trace is
soaked up by the fermion zero modes associated with the hair, the
partition function $Z^{hor}_{5D}$ measures the Witten index
$Tr(-1)^F=Tr(-1)^{2h+J}$
of the black hole microstates associated with the horizon
in a given $(n,Q_1,J)$ sector. 

\sectiono{Analysis of the four dimensional black hole entropy}
\label{s3}

Now we turn to the degeneracies of four dimensional black holes
obtained by placing the five dimensional black hole described above
at the center of Taub-NUT space. The corresponding solution is
given by\cite{0209114}
\ben \label{ep6}
dS^2 &=&  \left( 1 + { r_0   \over  r}\right)^{-1} 
\left[ - dt^2 +(dx^5)^2 + {r_0   \over r } (dt +
dx^5)^2 \right. \nonumber \\
&&
\left. +  
{\wJ \over 4}\left({1\over r}+ {4\over  R_4^{2}}\right)
 \,
(dx^4 + \cos\theta \, d\phi) 
\, (dt+ dx^5)\right] \nonumber \\
&&
 +\wh g_{mn}\,
 du^m du^n 
+ \left( 1 + { r_0   \over r }\right) \, ds_{TN}^2 \, , \nonumber \\
e^\Phi&=&\lambda \, , \nonumber \\
F^{(3)} &=& {r_0\over \lambda}\,  
\left(\epsilon_3 + *_6 \epsilon_3
+{1\over r_0} \left( 1 + { r_0   \over  r}\right)^{-1}
(dx^5+dt)\wedge d \wt\zeta \right)\, ,
\een
where
\be \label{edefzetanew}
\wt\zeta = -{\wJ \over 8} \left({1\over r} 
+{4\over R_4^{2}}\right) \, (dx^4 +\cos\theta d\phi)\, ,
\ee
\be \label{ep7}
ds_{TN}^2 =  \left({4\over R_4^{2}}+{1\over r}\right)^{-1} 
(dx^4 +\cos\theta d\phi)^2 + \left({4\over R_4^{2}}
+{1\over r}\right) \,
(dr^2 + r^2 d\theta^2 + r^2\, \sin^2\theta \, d\phi^2)\, .
\ee
Here $R_4$ is a constant labelling the asymptotic radius of the $x^4$
circle. Note that for $R_4^{ 2}=4r_0$ 
the 44, 45 and 55 components
of the metric become constant independent of $r$.
Thus $4r_0$ is the attractor value of $R_4^{ 2}$. We shall
proceed with the solution for general $R_4$.
Using \refb{ep7} we can express the solution 
given in \refb{ep6}
as
\ben \label{enewform}
dS^2 &=& - e_0^2 + e_1^2 + e_2^2 + e_3^2+e_4^2 + e_5^2
+\wh g_{mn} \, du^m\, du^n\, ,\nonumber \\
F^{(3)} &=& {r_0 \over \lambda\, r^2}\, 
\left[ \left(1+{r_0\over r}\right)^{-3/2}\,
\left({1\over r} +{4\over R_4^{2}}\right)^{-1/2}\, (e_2\wedge e_4\wedge e_5
+ e_0\wedge e_1\wedge e_3) \right. \nonumber \\ && 
\left.
+ {\wJ \over 8\, r_0} \, \left(1 +{r_0\over r}\right)^{-2}\,  
(-e_0\wedge e_2\wedge e_3 + e_0\wedge e_4\wedge e_5
-e_1\wedge e_2\wedge e_3 + e_1\wedge e_4\wedge e_5)
\right]\, , \nonumber \\
\een
where
\ben \label{eoneform}
e_0 &=& \left(1 +{r_0\over r}\right)^{-1} (dt+\wt\zeta), \nonumber \\
e_1 &=& \left(dx^5+dt- \left(1 +{r_0\over r}\right)^{-1} 
(dt+\wt\zeta) \right),   \nonumber \\ 
e_2 &=&
\left(1 +{r_0\over r}\right)^{1/2} \left({1\over r} 
+{4\over R_4^{2}}\right)^{-1/2} (dx^4 +\cos\theta d\phi), 
\nonumber \\ 
e_3 &=&
\left(1 +{r_0\over r}\right)^{1/2} \left({1\over r} 
+{4\over R_4^{2}}\right)^{1/2} \, dr
\, , \nonumber \\
e_4 &=&  \left(1 +{r_0\over r}\right)^{1/2} \left({1\over r} 
+{4\over R_4^{2}}\right)^{1/2} \, r\, d\theta
\, , \nonumber \\
e_5 &=& \left(1 +{r_0\over r}\right)^{1/2} \left({1\over r} 
+{4\over R_4^{2}}\right)^{1/2} \, r\, \sin\theta\, d\phi\, .
\een
Since $x^4$ has period $4\pi$, 
the asymptotic circle
parametrized by $x^4$ has finite radius.  
Thus asymptotically we have four
non-compact space-time dimensions. 
Also since $x^4$ now represents a compact coordinate, 
the quantum
number $J$ is
interpreted as the momentum along $x^4$ instead of angular
momentum.
However for small $r$ the
solution approaches that given in \refb{ep4}, and both solutions have
identical near horizon geometry.  To see this explicitly we take the
near horizon limit by first defining new coordinates $(\rho,\tau,y)$ via
\be \label{ep8}
r= r_0\, \beta\rho, \quad t= \tau/\beta, 
\quad  x^5=y-t
\ee
and taking the limit $\beta\to 0$. In this limit both 
\refb{ep4} and \refb{ep6} take the form\footnote{We could take
a more careful limit by beginning with a non-extremal black hole
and scaling the non-extremality parameter also by $\beta$
as reviewed in \cite{0809.3304}. However this does not
play any role in the present discussion.}
\ben \label{ep9}
dS^2&=&r_0 {d\rho^2\over \rho^2} + dy^2 
+r_0(dx^4 + \cos\theta d\phi)^2
+{\wJ \over 4 r_0} dy (dx^4 + \cos\theta d\phi) 
- 2\rho dy d\tau
 \nonumber \\ &&  +r_0  \left(
 d\theta^2 + \sin^2\theta d\phi^2\right) 
+ \wh g_{mn} du^m du^n\, , \nonumber \\
e^{\Phi} &=& \lambda\, , \nonumber \\
F^{(3)} &=& {r_0\over \lambda}\, \left[\eps_3 + *\eps_3
+{\wJ \over 8\, r_0^2}\, dy\wedge \left({1\over \rho}\, d\rho\wedge
(d \, x^4 +\cos\theta\, d\phi)
+ \sin\theta\, d\theta\wedge d\phi\right)\right]\, .
\een
Thus we expect that the contribution to the degeneracy from the
horizon degrees of freedom will be identical  for the four and
the five dimensional black holes. In particular 
the quantum entropy function will give identical results for
the two solutions.
We shall now try to test this at the microscopic level
by computing the degeneracies
associated with the four dimensional black hole horizon.

The microscopic degeneracy
associated with the four dimensional black hole is different from that
of the five dimensional black hole, as it receives additional
contribution from the modes living on the Taub-NUT space
as well as the modes associated with the motion of the D1-D5-brane
in the Taub-NUT space\cite{0605210}. 
If we denote by $d_{4D}(n,Q_1,J)$
the sixth helicity trace\footnote{$h$ now denotes the third component
of the angular  momentum in the (3+1) dimensional theory.
$J$ represents a U(1) charge in the four dimensional theory and
its inclusion in the trace is purely a matter of convenience.}
$-B_6\equiv -Tr((-1)^{2h+J}(2h)^6)/6!$
for the states of the four dimensional
black hole carrying quantum numbers
$(n,Q_1,J)$ then the four dimensional partition function defined via
\be \label{ep1a}
Z_{4D}(\rho,\sigma,v) = \sum_{n,Q_1,J} \, d_{4D}(n-1,Q_1, J)\, 
\exp\left[2\pi i \{(Q_1-1) \,  \sigma + (n-1) \,  \rho + J v\}\right]\, ,
\ee  
has
the 
form\cite{9607026,0412287,0505094,0506249,0605210}\footnote{The
correct sign of the partition function has been determined in
\cite{0708.1270}. Note that $d_{4D}(n,Q_1,J)$ used here differ from
the index used in \cite{0708.1270} by a factor of $(-1)^J$ due to the
insertion of $(-1)^J$ in our definition of $B_6$. However the 
definition of partition
function in \cite{0708.1270} has an explicit factor of $(-1)^{J+1}$
inserted.}
\be\label{ep2a}
Z_{4D}(\rho,\sigma,v) =
-e^{-2\pi i  
\rho -2\pi i\sigma -2\pi i v
 }\prod_{k,l,j\in \zzz 
\atop k,l\ge 0, j<0 \, for\, k=l=0}
\left( 1 - e^{2\pi i ( \sigma k   +  \rho l + v j)}
\right)^{-c(4lk - j^2)}\, .
\ee
Note that we now have $(n-1)$ in the argument
of $d_{4D}$ 
in \refb{ep1a}, matching the coefficient of $\rho$ in the exponent. 
This
reflects the fact that for the
four dimensional black holes the charge measured
at the horizon agrees with the charge measured by an asymptotic
observer. The $e^{-2\pi i\rho}$ factor in \refb{ep2a} 
is a reflection of the fact that the ground state of the Taub-NUT space
carries $-1$ unit of momentum along $S^1$; however this is visible 
only after taking into account the higher derivative term in the action
involving the gravitational Chern-Simons term. Finally we note that
there is no need to subtract the $n=0$ contribution from the sum, since
in the presence of a Taub-NUT space even the $n=0$ states are
quarter BPS. The near horizon geometry of the $n=0$ black hole will
however lose the memory of the Taub-NUT background and will have
enhanced supersymmetries.

We now need to remove the contribution to $Z_{4D}$
from the degrees of freedom living outside the horizon.
We begin by counting the 
fermionic modes living outside the horizon. 
First of all, there are 12 broken supersymmetry
generators leading to 12 fermion zero modes. 
They carry $h=\pm{1\over 2}$ and
soak up the $-(2\, h)^6/6!$ factor 
from the helicity trace. Thus
the effect of removing their 
contribution is to map the helicity trace index to the Witten
index of the remaining system\cite{0605210,0708.1270}. 
Had the black hole world-volume theory been Lorentz
invariant in the $(x^5,t)$ coordinates, 
eight of the zero modes would lift to right-moving fermion fields
and four of the zero modes would 
lift to left-moving fermion 
fields on the
black hole world-volume. 
As in the case of five dimensional black holes, 
we expect that
the breaking of Lorentz invariance does not affect the equations
for the left-moving modes and hence we should be able to lift the 
four left-chiral fermion zero modes
into full fledged left-moving  fermion
fields on
the black hole world-volume.
These modes produce a contribution to the
Witten index of the form
\be \label{eyy1}
\prod_{l=1}^\infty (1 - e^{2\pi i l \rho})^4\, .
\ee

Next we turn to the bosonic modes living on the black hole. 
As before we shall
proceed by pretending that the black hole world-volume has Lorentz
invariance in the $(x^5,t)$ plane, and then take into account the
lack of Lorentz invariance by freezing the right-moving fields.
Our arguments will be heuristic, but we give more explicit
construction of some of the modes in appendix \ref{sb}.
The black hole solution given in \refb{ep6} admits a
normalizable closed
2-form inherited from the normalizable
harmonic 2-form of the Taub-NUT space\cite{brill,pope}. 
It is given by
\be \label{eom1}
\omega = - 
{r \over 4r+ R_4^{ 2} } \, 
\sin\theta d\theta\wedge d\phi 
+{R_4^{ 2} \over (4r+ R_4^{ 2})^2} \,
 dr\wedge
(dx^4 + \cos\theta d\phi)\, .
\ee
Using the metric \refb{ep6} one can easily check that this harmonic
form is supported outside the near horizon throat geometry. 
Thus any 2-form field along this
harmonic form will give rise to a scalar mode living outside the
horizon. From the NSNS and RR 2-form fields of type IIB
string theory we get two scalar modes. Furthermore the 4-form field
with self-dual field strength, reduced on the 22 internal cycles of K3,
generate 3 right chiral and 19 left chiral 2-form
fields in type IIB string
theory on K3.\footnote{In our convention the right-chiral
2-form fields have self-dual 3-form field strength and the left-chiral
2-form fields have anti-self-dual 3-form field strength in six
dimensions.}
Picking up the components of these fields along
the 2-form $\omega$ we get 19 left-moving scalars and
3 right-moving scalars on the black hole world-volume.
By the logic given earlier we expect the left-moving modes to
survive even after taking into account the breaking of the Lorentz
invariance in the $(x^5-t)$ plane.
Besides these there 
are three goldstone bosons associated with
the three broken translational symmetries. After freezing the 
right-moving modes we get three more
left-moving modes on the black
hole world-volume.
Thus we have altogether 2+19+3=24 left-moving
scalars living outside the horizon.\footnote{Explicit 
form of these deformations can be found in appendix \ref{sb}.}
Since they do not carry any $J$ quantum number (which now
corresponds to momentum along $x^4$), 
their contribution to the black hole partition function is
given by
\be \label{eyy2}
\prod_{l=1}^\infty (1 - e^{2\pi i l \rho})^{-24}\, .
\ee

We shall now argue that the four dimensonal solution carries
four more left-moving bosonic excitations living outside the
horizon and carrying $J$-charge $\pm 1$. Explicit construction
of these modes have been discussed in appendix \ref{sb}.
Physically these modes represent the motion of the D1-D5 system
relative to the Taub-NUT space. Normally if in a composite system
we try to displace one component relative to the other there will
be a drastic change in the near horizon geometry and we would not
expect such deformations to be described by modes living outside the
horizon. However since the Taub-NUT space is non-singular 
everywhere,
the near horizon geometry of a D1-D5-Taub-NUT
system is described by that of the D1-D5 system,
and hence moving the Taub-NUT space relative to the
D1-D5 system should not alter the near horizon geometry. Thus
such deformations should be described by modes living outside the
horizon.
Furthermore since the coordinates labelling the transverse
position of the D1-D5 system transform in the vector
representation of $SO(4)$, these modes should carry
$J=\pm 1$. By the standard argument based on the 
lack of Lorentz invariance in the $x^5-t$ plane, we expect the
right-moving modes to be frozen but the left-moving modes should
be freely excitable.
The contribution from these modes to the partition function
is given by
\be \label{ezz1}
\prod_{l=1}^\infty \, \left[\left(1 - e^{2\pi i (l\rho+v)}\right)^{-2}
\left(1 - e^{2\pi i (l\rho-v)}\right)^{-2}\right]\, .
\ee

Can there be additional zero modes associated with the motion of
the D1-D5-system relative to the Taub-NUT space?
The five dimensional black hole world
volume in flat transverse space 
has four left-chiral fermion zero modes with
$(J,2h)=(0,\pm 1)$ and eight right-chiral 
fermion zero modes with $(J,2h)
=(\pm 1, 0)$, -- all 
living outside the horizon. By an argument similar to the one
in the previous paragraph, we expect them to be approximate
zero modes even when we place the five dimensional
black hole in the Taub-NUT
background. The four left-chiral fermion zero modes form part of the
12 goldstino zero modes of the combined system and have already
been counted before. Four of the eight right chiral fermion
zero modes must
form superpartners of the bosonic zero modes describing the 
motion of the D1-D5-brane system in transverse space. This gives
rise to a factor of $-e^{-2\pi iv} (1-e^{-2\pi i v})^{-2}$ from summing
over bound states in the supersymmetric quantum mechanics
describing the zero mode 
dynamics\cite{pope,9912082,0605210,0708.1270}. 
The other four right-chiral fermion zero
modes which are not paired with the bosons under supersymmetry
would give a factor of $-(e^{\pi i v} - e^{-\pi i v})^2$ since
they carry $J=\pm 1$. Thus these two factors cancel exactly and we
do not get any additional contribution to the hair
from these zero modes.

Combining \refb{eyy1}, \refb{eyy2} and \refb{ezz1}
we get the net contribution to the four dimensional black hole 
partition function from the hair:
\ben \label{eh1}
Z_{4D}^{hair}(\rho,\sigma,v) &=& \prod_{l=1}^\infty 
\left[\left(1 - e^{2\pi i l \rho}\right)^{-20}
\left(1 - e^{2\pi i (l \rho+v)}\right)^{-2}
\left(1 - e^{2\pi i (l \rho-v)}\right)^{-2}\right]\, .
\een

Let $Z_{4D}^{hor}$ denote the partition function of the horizon
degrees of freedom of the four dimensional black hole. Then naively
we have the relation $Z_{4D}=Z_{4D}^{hor}\times 
Z_{4D}^{hair}$, but as in the case of five dimensional black holes,
$Z_{4D}$ receives an 
extra contribution from the configuration where a small five dimensional
black hole carrying no momentum along $S^1$ is placed at the
center of the Taub-NUT space and the momentum along $S^1$ is
carried by the hair degrees of freedom. Denoting the extra 
contribution by $Z_{4D}^{extra}$ we have
\be \label{ecc1}
Z_{4D} = Z_{4D}^{hor}\times 
Z_{4D}^{hair} +Z_{4D}^{extra}\, .
\ee
$Z_{4D}^{extra}$ is given by the product of horizon partition
function of the small black hole as given in \refb{ecor4} and the
contribution from the hair degrees of freedom. 
The latter now consists of
four bosons and four
left- and four right-moving fermions associated with the motion
of the small black hole in Taub-NUT space, and eight right-moving
fermions, eight right-movimg bosons and twenty four left-moving
bosons associated with the fluctuations in Taub-NUT space.
Instead of going through a 
detailed analysis of these modes we simply note that
the number and dynamics of these modes is identical to those
describing the dynamics of the Taub-NUT space and the overall motion
of the D1-D5 system in Taub-NUT space as discussed in
\cite{0605210,0708.1270}. 
Thus the partition function associated with the hair
degrees of freedom can be read out from \cite{0605210,0708.1270}.
In particular the contribution 
from the degrees of freedom associated with the overall motion of
the D1-D5 system
can be read out from eq.(5.2.22) of \cite{0708.1270} 
for $N=1$:\footnote{The factor of 
$- e^{-2\pi i v} \, (1- e^{-2\pi i v})^{-2}$ arises from the sum over
bound states of the quantum mechanics describing the motion of the
D1-D5-system in Taub-NUT space. The main difference from the
computation of $Z_{4D}^{hair}$ is that when
the core of the black hole describing the D1-D5 system
carries zero momentum, we have only eight fermion zero modes
living on the D1-D5 system instead of twelve. Thus an extra factor
of $-(e^{\pi i v} - e^{-\pi i v})^2$ is missing here.}
\be \label{ecc2}
- e^{-2\pi i v} \, (1- e^{-2\pi i v})^{-2} \, 
\prod_{l\ge 1} (1-e^{2\pi i(l\rho+v)})^{-2} \, 
(1-e^{2\pi i(l\rho-v)})^{-2} \, (1-e^{2\pi il\rho})^4 \, .
\ee
On the other hand the
degrees of freedom of the Taub-NUT space contributes
\be \label{ecc3}
\prod_{l\ge 1} (1 - e^{2\pi i l\rho})^{-24}\, .
\ee
Taking the product of \refb{ecor4}, \refb{ecc2} and \refb{ecc3} gives
\ben \label{eadd1}
Z_{4D}^{extra}(\rho,\sigma,v) &=& - e^{-2\pi i (v+\rho+\sigma)} 
\, \left(1
-  e^{-2\pi i v}\right)^{-2} 
\, \, \prod_{k, j\in \zzz 
\atop k\ge 1}
\left( 1 - e^{2\pi i ( \sigma k   + v j)}
\right)^{-c(  - j^2)} \nonumber \\
&& \prod_{l=1}^\infty 
\left[\left(1 - e^{2\pi i l \rho}\right)^{-20}
\left(1 - e^{2\pi i (l \rho+v)}\right)^{-2}
\left(1 - e^{2\pi i (l \rho-v)}\right)^{-2}\right]\, .
\een
Using \refb{ep2a}, \refb{eh1}, \refb{ecc1} and \refb{eadd1},
and  the relations  
\be \label{ereln}
c(0)=20, \qquad c(-1)=2, \qquad c(u)=0 \quad 
\hbox{for $u\le -2$}\, ,
\ee 
we get
\ben \label{eadd2}
Z_{4D}^{hor}(\rho,\sigma,v) &=& (Z_{4D}-Z_{extra})
/Z_{4D}^{hair} \nonumber \\
&=& -e^{-2\pi i  
\rho-2\pi i \sigma
 }\, (e^{\pi i v}-e^{-\pi i v})^{-2}\,
 \prod_{k,l,j\in \zzz 
\atop k\ge 1, l\ge 0}
\left( 1 - e^{2\pi i ( \sigma k   +  \rho l + v j)}
\right)^{-c(4lk - j^2)}\nonumber \\
&& +  e^{-2\pi i  
\rho-2\pi i \sigma
 }\, (e^{\pi i v}-e^{-\pi i v})^{-2} \, \prod_{k, j\in \zzz 
\atop k\ge 1}
\left( 1 - e^{2\pi i ( \sigma k   + v j)}
\right)^{-c(  - j^2)}  \, .
\een
This is identical to
$Z_{5D}^{hor}$ given in
\refb{ese2}.
We also note that since the $-(2\, h)^6/6!$ 
term in the trace has been
absorbed by the fermion zero modes living outside the horizon,
$Z_{4D}^{hor}$ measures the Witten index $Tr(-1)^F$ of the
microstates associated with the horizon in a given
$(n,Q_1,J)$ sector. The equality of
$Z_{4D}^{hor}$ and $Z_{5D}^{hor}$ now
shows that the Witten indices associated with the near 
horizon degrees of freedom of the four and the five 
dimensional black holes are exactly identical.

Note added: It has  been shown in \cite{0907.0593} 
that the hair modes describing the transverse
oscillations of the five dimensional black hole, and the oscillations
of the BMPV black hole relative to the Taub-NUT space for the four
dimensional black hole, develop curvature singularities at the
future horizon. Thus they should not be included among the hair degrees
of freedom. Since they contributed the same amount to the respective
partition functions, the agreement between the partition functions of
four and five dimensional black holes after hair removal continue to
hold.

\medskip

{\bf Acknowledgement:} We would like to thank Gabriel Lopes Cardoso, Alejandra
Castro, Justin David, Bernard de Wit, Suvankar 
Dutta, Rajesh Gopakumar, Dileep
Jatkar, Swapna Mahapatra, Samir Mathur and Shiraz Minwalla for useful
discussions. NB would also like to thank the organizers of the YRC conference,
Perimeter, for  hospitality at the final stage of this work.
 
\appendix
 
\sectiono{Explicit construction of the left-moving bosonic
modes on the BMPV black hole} \label{sa}

Since our argument leading to the existence of left-moving modes
on the BMPV black hole
has been somewhat abstract we shall now explicitly demonstrate the
existence of such modes. For simplicity we shall 
%consider the $J=0$ solution and 
focus on the left-moving bosonic zero modes associated
with the transverse oscillations. If we introduce new coordinates
\ben \label{ecart1}
w^1 = 2\sqrt r \cos{\theta\over 2} \cos{x^4+\phi\over 2},
&& w^2 = 2\sqrt r \cos{\theta\over 2} \sin{x^4+\phi\over 2},
\nonumber \\
w^3 = 2\sqrt r \sin{\theta\over 2} \cos{x^4-\phi\over 2},
&& w^4 = 2\sqrt r \sin{\theta\over 2} \sin{x^4-\phi\over 2}\, ,
\een
then the solution given in \refb{ep4} takes the form
\ben \label{ep4new}
dS^2 &=& \psi(r)^{-1} 
 \left[ dx^+ dx^-+ (\psi(r)-1) (dx^+)^2  \right] + 
 \chi_i(\vec w)\, dx^+\, dw^i
 +  \wh g_{mn}\,
 du^m du^n 
 + \psi(r) \vec{dw}^2,
 \nonumber \\ 
 x^\pm &\equiv& x^5\pm t\, , \quad  
 r \equiv {1\over 4}\vec w^2, \quad
 \psi(r) \equiv \left( 1 + { r_0   \over  r}\right),
 \quad \chi_i(\vec w) \, dw^i = \psi(r)^{-1}\, {\wJ \over 4r}\,
 (dx^4+\cos\theta \, d\phi) \nonumber \\
 C^{(2)} &=& {1\over 2} C_{ij}(\vec w) dw^i\wedge dw^j +
 C_{+i}(\vec w) dx^+\wedge dw^i +
C_{+-}(\vec w) dx^+\wedge dx^-\, ,
 \een
where $C^{(2)}$ denotes the RR 2-form field and
$C_{ij}$, $C_{+i}$ 
and $C_{+-}$ are some fixed functions of $\vec w$.
We can now use the following algorithm to generate the
deformations  describing left-moving  
transverse oscillations of the black hole:
\begin{enumerate}
\item We first consider a deformation of the solution generated
by the diffeomorphism
\ben \label{edm1}
w^i &\to& w^i + a^i\, (x^++c)\, f + (x^++c) 
\, \vec a\cdot \vec w\, w^i \, g, \nonumber \\
x^- &\to& x^- - 2\, \vec a\cdot \vec w\, \psi^2\, f
- (x^++c) \,  \psi\, \left( \vec a\cdot \vec \chi\, f +
\vec a\cdot \vec w\, \vec w\cdot\vec \chi\, g
 \right), \nonumber \\
  x^+&\to& x^+\, ,
 \een
where $\vec a$ denotes an arbitrary constant four dimensional 
vector, $\vec a\cdot \vec w\equiv a^i w^i$,
$c$ is an arbitrary constant
and $f$ and $g$ are functions of $r$ satisfying
\be  \label{edeform2}
g = {1\over 2} \, \psi^{-2} \, (\psi^2 f)'\, .
\ee
Here $'$ denotes
derivative with respect to $r$.
The diffeomorphism has been chosen such that all the terms
in $\delta (dS^2)$ to first order in $a^i$
are proportional to $(x^++c)$ without any derivative acting on
it. By accompanying this diffeomorphism by a suitable gauge 
transformation of $C^{(2)}$ we can ensure that 
$\delta C^{(2)}$ also is proportional to $(x^++c)$ without any
derivative acting on it.
\item We now replace the overall factor of 
$x^++c$ by an arbitrary function
$\eps(x^+)$ everywhere in the deformed solution. Thus the deformed
configuration is proportional to $\eps(x^+)$. Furthermore, by construction
it is guaranteed to be a solution to the equations
of motion for $\eps(x^+)=x^++c$. 
This in turn shows that if we substitute the deformed configuration
into the
equations of motion then the terms proportional to
$\eps(x^+)$ and $\p_+\eps(x^+)$ must vanish
automatically. 
\item 
Our goal is to ensure that the deformed configuration is a
solution to the equations of motion to linear order in $\eps$ for
arbitrary function $\eps(x^+)$.
Since the field equations are second order in derivatives,
and terms involving $\eps(x^+)$ and $\p_+\eps(x^+)$ are
guaranteed to vanish,
it only remains to ensure that the terms involving $\p_+^2\eps$
vanish. Such terms can arise in the $++$ component of
the metric equation, and the vanishing of the term proportional
to $\p_+^2\eps$ can be shown to 
require\footnote{Note that since the three form field strengths
contain at most a single derivative
of $\eps$, they do not directly contribute any term
proportional
to $\p_+^2\eps$ in the equations of motion.}
\be \label{egij}
G^{ij}\delta G_{ij}=0\, ,
\ee
where $i,j$ run over the four transverse spatial coordinates,
$G_{ij}$ is the background metric and $\delta G_{ij}$ denotes
the first order deformation of the metric.
This imposes one additional constraint on the functions $f$ and
$g$. Once this condition is satisfied we have a set of deformations
parametrized by four arbitrary function 
$a^i \eps(x^+)$.\footnote{This analysis is 
similar in spirit, although much
simpler than, the one carried out in \cite{0712.2456}.}
\end{enumerate}
At the end of the second step this procedure 
gives
\ben \label{edeform1}
\delta \left(dS^2\right) &=& -{1\over 2}\, \eps(x^+)
\, \psi^{-2}\psi' \, \vec a\cdot \vec w 
\, (f + 4 r g) \, \left(dx^+ \, dx^- -(dx^+)^2\right) 
\nonumber \\ &&
+{1\over 2}
\, \eps(x^+) \, \psi' \,\vec a\cdot \vec w
\, (f+4rg) \, \vec{dw}^2  +  \eps(x^+)\, \psi \, f' \, 
\vec a\cdot \vec{dw} \, \vec w\cdot \vec{dw}
\nonumber \\
&&+ 2\, \eps(x^+)\, \psi \, g \, \vec a\cdot \vec{dw} \, \vec w\cdot \vec{dw}
+ 2\, \eps(x^+)\,\psi \, g \, \vec a\cdot \vec w  \, \vec{dw}^2 
 + \eps(x^+)\,
\psi \, g'\, \vec a\cdot \vec w\, 
\vec w\cdot\vec{dw} \, \vec w\cdot\vec{dw} \nonumber \\
&& - \eps(x^+)\, \psi^{-1} \, dx^+ \ d\left(\psi
(\vec a\cdot \vec\chi\, f + \vec a\cdot\vec w\, 
\vec w\cdot \vec \chi\, g)\right) \nonumber \\
&&
 +\eps(x^+) \, \chi_i\, 
 dx^+\, d\left(  a^i\, f +
 \vec a\cdot \vec w\,  w^i\, g
 \right) + \eps(x^+) \, \p_k \chi^i\, (a^k \, f + \vec a\cdot \vec w
 \, w^k\, g) \, dx^+ dw^i\, , \nonumber \\
\delta\, C^{(2)} &=& {1\over 2}\, \eps(x^+) \left(
\p_k \, C_{ij} + \p_i \, C_{jk} + \p_j \, C_{ki}\right)
\, (a^k f + 
\vec a\cdot \vec w\, w^k \, g) \, dw^i\wedge dw^j
\nonumber \\  
&& + \eps(x^+)\, \p_k C_{+-} \, (a^k f + 
\vec a\cdot \vec w\, w^k \, g) \, dx^+\wedge dx^-
\nonumber \\
&&
+ \eps(x^+)\, \p_k C_{+-} \, dw^k\wedge 
\, (2\, \psi^2 \, f\, \vec a\cdot \vec{dw}
+ \vec a\cdot \vec w\, (\psi^2 f)' \, \vec w\cdot \vec{dw})
\nonumber \\ &&
-\eps(x^+)\,  
\p_k \, C_{+-} \,  \left( \vec a\cdot \vec \chi\, f +
\vec a\cdot \vec w\, \vec w\cdot\vec \chi\, g
 \right)\, \psi\, dw^k\wedge dx^+ \nonumber \\
 && + \eps(x^+)\, (\p_l C_{+k} -\p_k C_{+l})\, (
 a^l\,  f +  \vec a\cdot \vec w\, w^l \, g)
  \, dx^+\wedge dw^k\, .
\een
Substituting this into \refb{egij}
gives
\be \label{edeform3}
2\psi' \, (f+4rg) + \psi \, f' + 10\, \psi \, g + 4 \, r \, \psi 
\, g' = 0\, .
\ee
Using eq.\refb{edeform2} we can regard \refb{edeform3}
as a second order linear differential equation for $f$. Thus it
has two independent solutions. 
It is easy to verify that the general solution to \refb{edeform2},
\refb{edeform3} is
\be \label{egensol}
f = (A_0\, r^{-2}+B_0)\, \psi^{-2}, \qquad g = - A_0\, r^{-3}\,
\psi^{-2}\, ,
\ee
where $A_0$ and $B_0$ are two arbitrary constants. Requiring that
the solution gives a normalizable deformation of the metric and
the 2-form field near $r=0$ we get $A_0=0$. Thus we have
\be \label{esol2}
f = B_0\, \psi^{-2}\, , \qquad g = 0\, .
\ee
It is easy to verify that the deformations of the metric and the
2-form field associated with this choice of $f$ is normalizable both
at $r=0$ and at $r=\infty$.
Thus we have normalizable deformation of
the solution parametrized by four indendent functions $a^i\,
\eps(x^+)$. This
shows the existence of four left-moving modes on the 
black hole
world-volume. Furthermore the contribution to the
norm of the deformation from  the throat region $r<<r_0$ vanishes,
showing that these modes are located outside the horizon.

We expect that a similar argument can be used to construct the
four left-moving fermionic modes on the black hole world-volume.
In this case we shall need to use the broken supersymmetry
generators to generate the fermionic deformation of the solution.
However we shall not carry out this analysis explicitly.

\sectiono{Explicit construction of the
left-moving bosonic modes
on the four dimensional black hole} \label{sb}

In this appendix we shall give explicit construction of the bosonic
zero modes living on the four dimensional black hole.
We begin with the left-moving
zero modes associated with the harmonic two form
$\omega$ in the Taub-NUT space given in \refb{eom1}.
For any 2-form field $B$ -- either
the NSNS or RR sector 2-form field of the ten dimensional type IIB
string theory or a four form field with two legs along an internal 2-cycle
of $K3$ -- we consider a deformation of the form 
\be \label{eagaina}
\delta B = \eps(x^+) \, \omega\, ,
\ee
for any function $\eps(x^+)$ of $x^+=x^5+t$. 
This gives
\ben \label{eagainb}
d(\delta \, B) &=&
\eps'(x^+)\, dx^+\wedge\omega\nonumber \\
&=& -\eps'(x^+)\,  
{1\over r^2\, R_4^2}\, \left({1\over r} +{4\over R_4^2}\right)^{-2}\,
\left(1+{r_0\over r}\right)^{-1}\, \nonumber \\ &&
(e_0\wedge e_2\wedge e_3 + e_0\wedge
e_4\wedge e_5 + e_1\wedge e_2\wedge e_3 + e_1\wedge
e_4\wedge e_5)\, ,
\een
where the 1-forms $e_i$'s have been defined in \refb{eoneform}.
$d(\delta B)$ given in \refb{eagainb} can be shown to be 
anti-self-dual. Hence $d(\delta B)$ is both closed
and co-closed and $\delta B$ given in \refb{eagaina}
provides a solution to the linearized equations of motion 
of $B_{\mu\nu}$ around the
background \refb{ep6}.
For the  3-form field strength deformation given in
\refb{eagainb} one also finds 
that there is no contribution to the stress tensor from the
interference term between the deformation and the leading
order field strength given in \refb{enewform}. 
As a result the deformation \refb{eagaina}
also satisfies the metric equation of motion at the linearized level.
However in order that \refb{eagaina} corresponds
to a valid configuration in string theory, $B$ must
correspond to a left-chiral 2-form (which has anti-self-dual
field strength
in our convention). Since type IIB on K3 has 2+19=21 left-chiral
2-form fields we get 21 left-moving bosonic  modes from this
construction.
Finally this deformation is
normalizable with the metric given in \refb{ep6} and the norm is
supported outside the throat, \i.e.\ outside the $r<<r_0, R_4^{ 2}$
region. Thus these modes should be counted as part of the
black hole hair.

Next we shall describe the left-moving modes associated with
the 3 transverse motion of the black hole. 
For this we introduce new coordinates $(y^1,y^2,y^3)$ via
\be \label{eydef}
y^1 = r \, \cos\theta\, \cos\phi, \qquad y^2 = r \, \cos\theta\,
\sin\phi, \qquad y^3=r\, \cos\theta\, .
\ee
In this coordinate system the metric given in eq.\refb{ep6}
takes the form
\ben \label{enf1}
dS^2 &=& \psi(r)^{-1}\, \left\{dx^+ dx^- + 
(\psi(r)-1)\, (dx^+)^2\right\} +{\wJ \over 4}\, \chi(r)\, \psi(r)^{-1}
\, (dx^4 + A_\alpha (\vec y)\, dy^\alpha)\, dx^+ \nonumber \\
&& + \psi(r) \, \chi(r)^{-1} \, \left(dx^4 + A_\alpha (\vec y)\, 
dy^\alpha\right)^2
+ \psi(r) \, \chi(r) \, \vec {d y}^2 + \wh g_{mn}\, du^m du^n\, ,
\een
where
\be \label{enf2}
\psi(r) = 1 +{r_0\over r}\, , \quad \chi(r) = {1\over r}
+{4\over R_4^{ 2}}\, , \quad A_\alpha(\vec y) \, dy^\alpha
= \cos\theta\, d\phi\, .
\ee
We can now generate an $x^+$ dependent deformation of
this solution by first considering a diffeomorphism
\ben \label{enf3}
y^\alpha &\to& y^\alpha + (x^++c)\, (b^\alpha \, \wt f + \vec b\cdot \vec y\, y^\alpha\,
\wt g)\, , \nonumber \\
x^- &\to & x^- - 2\, \vec b\cdot \vec y\, \chi\, 
\psi^2\, \wt f\, , \nonumber \\
x^4 &\to& x^4 - (x^++c)\, A_\alpha\, (b^\alpha \, \wt f 
+ \vec b\cdot \vec y\, y^\alpha\,
\wt g)\, ,
\een
and then replacing $(x^++c)$ by $\eps(x^+)$ in the deformed
solution. 
Here $c,b^1,b^2,b^3$ are arbitrary parameters, $\vec b\cdot \vec y
\equiv b^\alpha\, y^\alpha$,
and $\wt f$ and $\wt g$
are functions satisfying
\be \label{enf4}
\wt g ={1\over r} \, \psi^{-2}\, \chi^{-1} (\psi^2\,\chi\, \wt f)'\, .
\ee
This gives
\ben \label{enf5}
\delta (dS^2) &=& -\eps(x^+)\, \psi^{-2} \, \psi'\, {\vec b\cdot 
\vec y \over r}\, (\tilde f+ r^2 \tilde g) \, \left( dx^+ dx^- - (dx^+)^2\right) 
\nonumber \\ &&
+{\wJ \over 4}\, \eps(x^+)\, (\psi^{-1}\chi)'
{\vec b\cdot 
\vec y \over r}\, (\tilde f+ r^2 \tilde g)  
\, dx^+ \, (dx^4 + \vec A\cdot \vec{dy})
\nonumber \\ &&
+ {\wJ \over 4}\, \eps(x^+)\, \psi^{-1}\, \chi\,
(\p_\alpha A_\beta - \p_\beta A_\alpha)\, (b^\alpha \, \wt f 
+ \vec b\cdot \vec y\, y^\alpha\,
\wt g) \, dy^\beta\, dx^+
\nonumber \\
&& + \eps(x^+)\, (\psi\, \chi^{-1})' \, {\vec b\cdot 
\vec y \over r}\, (\tilde f+ r^2 \tilde g)  
\, (dx^4 + \vec A\cdot \vec{dy})^2 \nonumber \\
&& + 2\, \eps(x^+)\, \psi\, \chi^{-1}\, 
(\p_\alpha A_\beta - \p_\beta A_\alpha)\, 
(b^\alpha \, \wt f + \vec b\cdot \vec y\, y^\alpha\,
\wt g) \, dy^\beta\, (dx^4 + \vec A\cdot \vec{dy})\nonumber \\
&& + \eps(x^+)\, (\psi\, \chi)'  
\, {\vec b\cdot 
\vec y \over r}\, (\tilde f+ r^2 \tilde g) 
\, \vec dy^2 + 2\, \eps(x^+)\, \psi\, \chi\, 
dy^\alpha \, d\left( b^\alpha \, \wt f + \vec b\cdot \vec y\, y^\alpha\,
\wt g \right)\, . \nonumber \\
\een
One can construct the deformation of the 2-form field in a
straightforward manner but we shall not do this here.\footnote{For
this one
needs to accompany the diffeomorphism \refb{enf3}
by an
appropriate gauge transformation of the 2-form field such that
every term in the deformation has an explicit factor of $(x^++c)$
without any derivative acting on it. 
We then replace $(x^++c)$ by $\eps(x^+)$.}
Our construction guarantees that when we substitute the deformation
\refb{enf5} (and the corresponding deformation of the 2-form
field) into the linearized equations of motion in the black hole
background, all terms up to first derivative of $\eps(x^+)$
vanish. Requiring the coefficient of the $\p_+^2 \eps$ term to
vanish gives us the equation:
\be \label{enf6}
\psi^{-1}\, \chi \, (\psi\, \chi^{-1})' (\wt f + r^2\, \wt g)
+ 3\, \psi^{-1}\, \chi^{-1} \, (\psi\, \chi)' (\wt f + r^2\, \wt g)
+ 2\, (\wt f' + 4\, r\, \wt g+ r^2 \, \wt g') = 0\, .
\ee
Using eq.\refb{enf4} we can regard \refb{enf6}
as a second order linear differential equation for $\wt f$. Thus it
has two independent solutions. 
It is easy to verify that the general solution to \refb{enf4},
\refb{enf6} is
\be \label{egensol2}
\wt f = (\wt A_0\, r^{-3}+\wt B_0)\, \psi^{-2}\, \chi^{-1}, 
\qquad \wt g = - 3\, \wt A_0\, r^{-5}\,
\psi^{-2}\, \chi^{-1}\, ,
\ee
where $\wt A_0$ and $\wt B_0$ are two 
arbitrary constants. Requiring that
the solution gives a normalizable deformation of the metric and
the 2-form field near $r=0$ we get $\wt A_0=0$. Thus we have
\be \label{esol2new}
\wt f = \wt B_0\, \psi^{-2}\,  \chi^{-1}, \qquad \wt g = 0\, .
\ee
It is easy to verify that the deformations of the metric and the
2-form field associated with this choice of $\wt f$ 
is normalizable both
at $r=0$ and at $r=\infty$.
Thus we have normalizable deformation of
the solution parametrized by three 
indendent functions $b^\alpha\,
\eps(x^+)$. This
shows the existence of three left-moving modes on the 
black hole
world-volume
describing the
left-moving transverse oscillation modes of the black hole. 
Furthermore the contribution to the
norm of the deformation from  the throat region $r<<r_0,R_4^2$ 
vanishes,
showing that these modes are located outside the horizon.

Finally we turn to the zero modes describing the motion of
the D1-D5 system relative to the Taub-NUT space.
We shall not carry out the construction in detail but describe these
deformations in the  limit $R_4^{ 2}>>r_0$.
To leading order in this limit, the deformations
associated with these left-moving modes
are in fact
given by the ones described in \refb{edeform1}. 
Indeed the arguments of appendix \ref{sa}
show that for $r<<R_4^{ 2}$ when the Taub-NUT metric
can be replaced by flat metric, the deformations given in
\refb{edeform1} satisfy the linearized equations
of motion. On the other hand since the function $f$ in
\refb{edeform1} approaches
a constant for $r>>r_0$, the metric fluctuations fall off as $1/r^2$
and the contribution to the norm of the deformation from this region
is small.
Thus the deformation given in \refb{edeform1}
is supported in the region $r\sim r_0$, and
for $r\sim R_4^{ 2} >> r_0$, where the deviation of the
Taub-NUT metric from the flat metric becomes significant, the
deformation is close to zero. Thus we conclude that
in the region where the
deformation \refb{edeform1} is supported it
remains an approximate
solution to the equations of motion.\footnote{While this argument
has been somewhat heuristic, we note that even in the microscopic
counting the transverse oscillation modes of the D1-D5 system
in Taub-NUT space was accounted for by assuming that for large
$R_4$ we can regard the non-zero mode oscillations of the
D1-D5 system as free oscillators\cite{0605210}.}

Our analysis also allows us to determine the $J$ quantum numbers
of various deformations.
Since in the region $r<< R_4^{ 2}$ the parameters
$\vec a$ labelling the deformation in \refb{edeform1}
transform in the vector 
representation
of the $SO(4)$ rotation group acting on the coordinates
$\vec w$, they carry $J=\pm 1$.
This may also be seen by noting that under a translation
$x^4\to x^4 + \beta$, these modes transform with a phase
$e^{\pm i \beta / 2}$. Since $x^4$ has period $4\pi$, this shows that
these modes carry $\pm 1$ quantum of $x^4$ momentum.
On the other hand the deformations describing the overall transverse
motion of the black hole, described by the parameters $b^\alpha$, are
neutral under $x^4$ translation, and hence has $J=0$.
The different transformation properties of the modes
labelled by $\vec a$ and $\vec b$  help demonstrate that
they are distinct deformations of the solution.

%\small
%\baselineskip 12pt
%\parskip -12pt

\end{document}